\documentclass[twocolumn,secnumarabic,amssymb,nobibnotes,aps,prl]{revtex4-1}

\usepackage{graphicx}
\usepackage{dcolumn}
\usepackage{bm}
\usepackage{multirow}
\usepackage{mathptmx}
\usepackage{mathtools} 
\usepackage{amsmath}
\usepackage{amssymb}
\usepackage{bibentry,natbib}
\usepackage{setspace}
\usepackage{mathrsfs}
\usepackage[linkcolor=blue,citecolor=blue]{hyperref}

\begin{document}
\title{Probing dissipation process via Fano resonance and collective effect in the X-ray cavity}
\author{Tian-Jun Li,$^{1}$ Xin-Chao Huang,$^{1,2}$ $^\dag$ Zi-Ru Ma,$^{1}$ Bo Li,$^{1}$ and Lin-Fan Zhu $^{1}$ $^\ast$}
\affiliation{$^1$ Hefei National Laboratory for Physical Sciences at Microscale and Department of Modern Physics, University of Science and Technology of China, Hefei, Anhui 230026, People's Republic of China.\\ E-mail: lfzhu@ustc.edu.cn\\
$^2$ Femtosecond X-ray Experiments Group, European XFEL, Holzkoppel 4, 22869 Schenefeld, Germany.\\ E-mail: xinchao.huang@xfel.eu\\}
\begin{abstract}
In the absence of time-reversal symmetry, the asymmetric parameter \emph{q} of the Fano resonance is extended into the complex space, where its trajectory can be utilized to investigate the decoherence process. By embedding the ensemble of M$\ddot{\rm{o}}$ssbauer nuclei in the thin-film planar cavity in this work, the trajectories of asymmetric parameter \emph{q} are studied via the collective effect of the nuclear ensemble, which is regulated by the nuclear abundance and angle offset. Due to the diverse controllable methods of the collective resonant strength, there are different straight lines and arc-shape trajectories in the complex plane, in which the slopes and the radius can be respectively adjusted by the angle offset and nuclear abundance. It is demonstrated that the dissipation process can be suppressed equivalently by the strong energy exchange between the cavity and nuclear ensemble.
The present results could enrich the behaviors of the asymmetric parameter $q$ in the complex plane and would provide new possibility for the decoherence research through the thin-film planar cavity.
\end{abstract}
\maketitle

\section{ Introduction}
Quantum-to-classical transition inaugurates the development of the decoherence theory, which, in a nutshell, states that quantum coherence in a system will be destroyed by the interaction with the environment \cite{Zurek-2003,Schlosshauer-2019}. The environment induced decoherence is a serious obstacle against the preservation of the long coherence times in a quantum information devices. As a result, decoherence piques academics' attentions as one of the most fascinating challenge topics in quantum physics.  Within the framework of Markov approximation, decoherence dynamics is conveniently characterized by Markovian master equation with Liouville superoperator in theory \cite{Schlosshauer-2019}, but it is still difficult to distinguish the different decoherence processes from the observables \cite{Rotter-2005}.

Fano resonance, a ubiquitous phenomenon known as its asymmetric profile, occurs when a discrete state interferes with a continuum one, and it has been studied in a wide array of different subfields of physics including neutron scattering \cite{Adair-1949},  Raman scattering \cite{Cerdeira-1973}, four-wave-mixing-induced autoionization \cite{Armstrong-1974}, single-electron transistor \cite{J-2001}, quantum dots \cite{Kobayashi-2002} and fast-electron impact \cite{LXJ-2003}, to name a few (for a review see \cite{Andrey-2010}). Generally, the asymmetric profiles of the Fano resonances are described by a well-known asymmetric parameter $q$ that is relevant to the phase difference of the two pathways \cite{connerade-1988,Ott-2013,Limonov-2017}. 
However, for an open quantum system the environment induced decoherence represents the increasing of the entropy, i.e. the broken of the time-reversal symmetry (TRS). In the absence of TRS, the asymmetric parameter \emph{q} of the Fano resonance takes on complex values \cite{Clerk-2001,Kobayashi-2002}. As a result, not only the specific decoherence type can be distinguished by the unique \emph{q} behavior, but also the degree of the decoherence can be presented by the position of the asymmetric parameter \emph{q} in the complex plane. For the Fano resonance only influenced by the dissipation process, the coupling to the environment leads to a wider resonant width, and the corresponding asymmetric parameter $q$ displays a straight line in the complex plane. As the dissipation increases, the asymmetric parameter $q$ will follow the straight line towards the extreme point $q=i$ \cite{Andreas-2010}. 

In the last decade, the development of high-brilliance synchrotron radiation and X-ray free electron laser (XFEL) has broadened the quantum optics research to the x-ray regime \cite{Adams-2013,Chumakov-2018,Schneider-2018}. Using a nuclear or inner-shell excitation of the atomic ensemble embedded in a thin-film planar cavity, a large number of fundamental phenomena have been studied based on this cavity setup, including the collective Lamb shift and superradiance \cite{Ralf-2010,Haber-2019,Huang-2021}, electromagnetically induced transparency \cite{Ralf-2012}, spontaneously generated coherence \cite{Heeg-2013-2}, X-ray storage \cite{Kong-2016}, Rabi oscillations \cite{Haber-2017} and etc. 
In addition, this thin-film planar cavity can also be used as a phase detection via the Fano resonance, where the phase difference between the two pathways is controlled by the cavity detuning conveniently \cite{Heeg-2015}. Note here that the asymmetric parameter $q$ in this platform can be a complex value due to the incoherence loss channel, which opens the door to study the dissipation process from the trajectory of the asymmetric parameter $q$ in the complex plane. 

In the present work, the dissipation process, affected by the collective resonant strength (CRS), is studied via the Fano resonances in the thin-film planar cavity, where the CRS can be regulated by the nuclear abundance and the angle offset respectively. Calculated by the benchmark CONUSS package \cite{conuss-2000}, the Fano resonances with changeable CRS are fitted and discussed using the phenomenological quantum model \cite{Heeg-2015-2,Heeg-2013}. It is found that the asymmetric parameter $q$ presents a straight line towards the extreme point $q=i$ in the complex plane as the nuclear abundance decreases, while it presents arc-shape trajectory with varying the angle offset. The present results show that the collective effect manipulated by the nuclear ensemble can not only suppress the dissipation process, but also significantly change the arc trajectory, which enrich the approaches to study the dissipation.
\section{ Theoretical Model} 
\begin{figure*}[htbp]
	\centering
	\includegraphics[width=\textwidth]{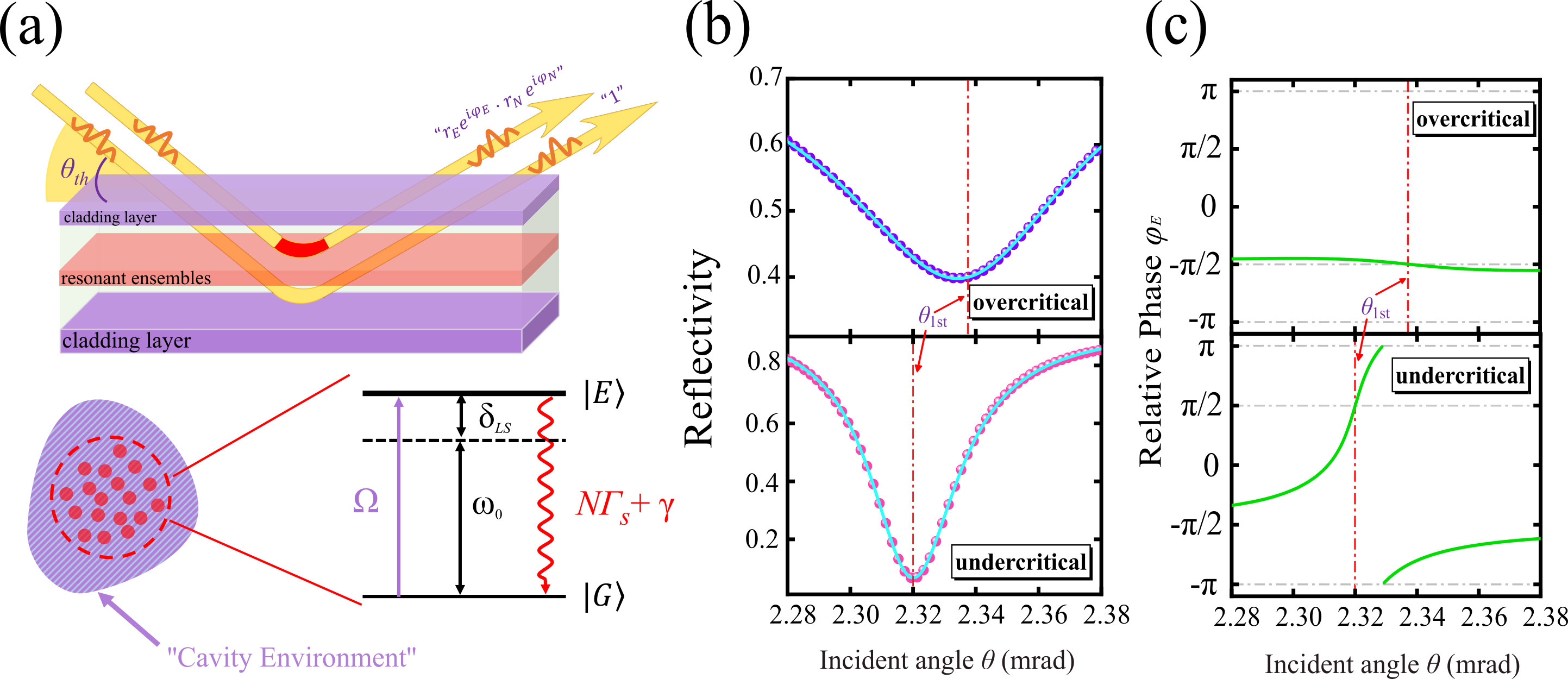}
	\renewcommand{\figurename}{Fig.}
	\caption{\label{Fano_full} (a) The schematic design of the thin-film planar cavity. The cladding layers (purple) are made by the same high density material of Pt for overcritical regime and Pd for undercritical regime, while the low density material of C was chosen to stack the cavity with the M$\rm{\ddot{o}}$ssbauer nuclei (pink) of $^{57}\rm{Fe}$ embedded in the middle. At the mode angle $\theta_{th}$, the cavity is driven by the external x-ray field with a large energy width ($\sim$ 100 eV) and the ensemble of nuclei is excited by the cavity mode and decays with the superradiance $N\Gamma_s+\gamma$ ($\sim \mu$eV ) \cite{Ralf-2010}. Fano resonance is formed between the two pathways of a continuum and a descrete state, in which the former is the cavity mode shown as ``1'', while the latter is the nuclear resonance ${r_N}{e^{i{\varphi _N}}}$ with the cavity environment ${r_E}{e^{i{\varphi _E}}}$. (b) The numerical reflectivities of the bare cavity caculated by the benchmark CONUSS package \cite{conuss-2000} (dots) along with their fitted curves by the quantum model \cite{Heeg-2013,Heeg-2015-2} in overcritical and undercritical regimes around the first cavity mode angle $\theta_{1st}$. (Details of the fitting processes are described in the Appendix) (c) The relative phase $\varphi_E$ in the overcritical and undercritical regimes in the vicinity of the first mode angle $\theta_{1st}$.}
\end{figure*}

The thin-film planar cavity is composed of seveal cladding layers and guiding layers as shown in Fig. \ref{Fano_full}(a). The cladding layers act as mirrors with high electron density materials such as Pd and Pt, while the guiding layers stack the cavity space with low electron density materials such as $\rm{B_4}$C, B and C (C is used in this work).  At some certain incident angles $\theta_{th}$, the cavity modes can be driven to generate various standing waves by the external x-ray field. When the resonant M$\ddot{\rm{o}}$ssbauer nuclei are doped in the center of this platform, the nuclear ensemble can be excited by a single photon to the collective state \cite{Scully-2009}
\begin{align}
\left| E \right\rangle {\rm{ = }}\frac{1}{{\sqrt N }}\sum\limits_{n = 1}^N {\left| {{g_1}} \right\rangle  \cdots } \left| {{e_n}} \right\rangle  \cdots \left| {{g_N}} \right\rangle\tag{1}
\end{align}
where $\left| {{g}} \right\rangle$ and $\left| {{e}} \right\rangle$ are the single atom ground and excited state. As a result, Fano interference occurs as shown in the Fig. \ref{Fano_full}(a), where the photon reflected by the cavity directly serves as the continuum, while the one interacting with the resonant nuclear ensemble is regarded as the discrete state \cite{Heeg-2015}. In addition,  the resonant energy of the nuclear ensemble can be shifted by the collective Lamb shift $\delta_{LS}$ and the line width can be broadened by the superradiance $N\Gamma_s+\gamma$ \cite{Ralf-2010,Scully-2010}. 

Considering the incoherent processes in the phenomenological quantum model developed in Refs. \cite{Heeg-2013,Heeg-2015-2}, the reflectivity we invoke here is given by 
\begin{align}
\left| R \right|^2 &= {\left| {{\frac{1}{{{r_E}}} \cdot \frac{{{2\kappa _R}}}{\kappa }}} \right|^2} \cdot {\left| {1 +  {r_E}{e^{i\varphi_{E}}} \cdot \frac{{\frac{{N{\Gamma _{\rm{s}}}}}{2}}}{{\omega  - {\omega _0} + {\delta _{LS}} + \frac{i}{2}(N{\Gamma _{\rm{s}}} + \gamma )}}} \right|^2}\nonumber\\
  &= {\left| {{\frac{1}{{{r_E}}} \cdot \frac{{{2\kappa _R}}}{\kappa }}} \right|^2} \cdot {\left| {1 + \Pi \; \cdot \; \frac{{{r_E}}}{{\varepsilon + i}}\;\cdot \; {{e^{i\varphi_{E}}}}} \right|^2}\tag{2}
\end{align}
Here, $\kappa$ is the cavity decay rate and $\kappa_R$ is the coupling strength of the external x-ray field into the cavity mode. $\varepsilon  = 2{{(\omega  - {\omega _0} + \delta_{LS} )} \mathord{\left/{\vphantom {{(\omega  - {\omega _0} + \delta_{LS} )} {(\gamma  + N{\Gamma _s})}}} \right.\kern-\nulldelimiterspace} {(\gamma  + N{\Gamma _s})}}$ is a dimensionless energy scaled by the  superradiance $N\Gamma_s + \gamma$ \cite{Ralf-2010,Heeg-2013,Huang-2021}, where $\Gamma_{S} = {{{2g^2}} \mathord{\left/{\vphantom {{{2g^2}} {(\kappa  + {{\Delta _c^2} \mathord{\left/{\vphantom {{\Delta _c^2} \kappa }} \right.\kern-\nulldelimiterspace} \kappa }}}} \right.\kern-\nulldelimiterspace} {(\kappa  + {{\Delta _c^2} \mathord{\left/{\vphantom {{\Delta _c^2} \kappa }} \right.\kern-\nulldelimiterspace} \kappa }}})$ represents the single atom broadening width with the cavity-nucleus coupling constant $g$. $\Pi  = {{N{\Gamma _s}} \mathord{\left/{\vphantom {{N{\Gamma _s}} {(N{\Gamma _s} + \gamma )}}} \right.\kern-\nulldelimiterspace} {(N{\Gamma _s} + \gamma )}}$ is the collective resonant strength with the values from 0 to 1. $r_E=\frac{{{2\kappa _R}}}{\kappa }\sqrt {\frac{{{\kappa ^2} + \Delta _c^2}}{{{{(2{\kappa _R} - \kappa )}^2} + \Delta _c^2}}}$ is the relative amplitude and ${\varphi _E}{\rm{ = }}\arg (\kappa  - i\Delta_c) + \arg (2{\kappa _R} - \kappa  + i\Delta_c) - {\pi  \mathord{\left/{\vphantom {\pi  2}} \right.\kern-\nulldelimiterspace} 2}$ is the relative phase of the two pathways. $\Delta_c$ is the energy detuning induced by the angle offset and can be written as
\begin{align}
{\Delta _c} = &\left( {\frac{{\sin {\theta _{th}}}}{{\sin \theta }} - 1} \right){\omega_0} \tag{3}
\end{align} 
where $\theta_{th}$ is the cavity mode angle and $\theta$ is the grazing incident angle. $\omega_0$ represents the energy of the nuclear excitation. Varying with the incident angle $\theta$, the relative phase $\varphi_{E}$ can be changed correspondingly. In the vicinity of the cavity mode, the cavity decay rate $\kappa$ and the coupling strength $\kappa_R$ can be regarded as constants, thus the relative amplitude and phase can be determined by the cavity detuning ${\Delta_c}$ in the thin-film planar cavity.

The relative phase $\varphi_E$ that depends on the cavity detuning $\Delta_c$ has remarkably different behaviors, determined by the relationship between the cavity decay rate $\kappa$ and the coupling strength $\kappa_R$. 
In detail, according to the ratio of the cavity decay rate $\kappa$ to the coupling strength $\kappa_R$, the cavity can be classified into three distinct regimes through adjusting the thickness of the top cladding layers or changing the materials \cite{Heeg-2013} viz. 
$$\left\{ \begin{array}{l}
{2{\kappa _R}} > \kappa, \; \rm{overcritical\; regime}\\
{2{\kappa _R}} = \kappa, \; \rm{critical\; regime}\\
{2{\kappa _R}} < \kappa, \; \rm{undercritical\; regime}
\end{array} \right.$$
In the critical regime, the amplitude $r_E$ and phase $\varphi_E$ make the complex \emph{q} approach to a real value \cite{Heeg-2015}, which corresponds to the usual Fano resonance, i.e., the asymmetric parameter $q$ locates at the real axis in the complex plane. In the overcritical and the undercritical regimes, although the asymmetric parameter $q$ is always a complex value, the behaviors of the relative phase $\varphi_E$ are different with varying the incident angle. 
To avoid the effect from the inhomogeneous ensembles \cite{Adriana-2021,You-2003} and considering a single polarization and unmagnetized case for brevity, the cavity in the overcritical regime is designed as Pt\,(0.5nm)\,/ C\,(20.8nm)\,/ $^{57}$Fe\,(0.3nm)\,/ C\,(19.6nm)\,/ Pt\,(2.5nm), and the undercritical one is Pd\,(3.5nm)\,/ C\,(20.8nm)\,/ $^{57}$Fe\,(0.3nm)\,/ C\,(19.6nm)\,/ Pd\,(2.5nm) in this work. Driven by the first cavity mode, the reflectivities of the bare cavities are fitted to obtain the cavity decay rate $\kappa$ and the coupling strength $\kappa_R$ as shown in Fig. \ref{Fano_full}(b), and their relative phases varying with the incident angle are shown in Fig. \ref{Fano_full}(c). It is apparent that the relative phase $\varphi_{E}$ around the cavity mode angle locates in a narrower range around $-\pi/2$ in the overcritical regime, while a wider range around $\pi/2$ is observed in the undercritical regime. 
 
As mentioned above, the resonant width of the Fano resonance can be influenced by the dissipation process. For the microwave cavity, the strong dissipation can broaden the resonant width, which is realized by choosing different materials or temperature \cite{Andreas-2010}. In the thin-film planar cavity, the strong dissipation corresponds to a very large energy width of the bare cavity ($\sim$ 100 eV), which makes such setup insensitive to the small variation of the dissipation. However, the collective coupling between the cavity and nuclear ensemble can affect the dissipation, and its effect can be detected by the asymmetric parameter $q$. Actually, the atomic ensemble was often uesd in the optical cavity to enhance the strength of the atom-photon coupling without requiring a high finesse cavity, such as the self-induced Rabi oscillations \cite{Haroche-1983} and trapping photon in the atomic media \cite{Lukin-2000}. Similarly, in this work the resonant width of Fano resonance will be broadened by the collective effect of the nuclear ensemble, where the CRS is relevant to the effective nuclear number $N$ and the single atom broadening width $\Gamma_{\rm{s}}$. For the certain cavity structure, the $\Gamma_{\rm{s}}$ is only relevant to the angle offset. As for the effective nuclear number $N$, it can be changed by the thickness or the abundance of the resonant layer. In order not to influence the cavity properties, the effective nuclear number is changed by the nuclear abundance in this work. Hence, the dissipation process with changeable collective effect can be studied by adjusting the nuclear abundance and angle offset respectively in such platform.

\section{Results and Discussion}
\begin{figure*}[!htbp]
	\centering
	\includegraphics[width=\textwidth]{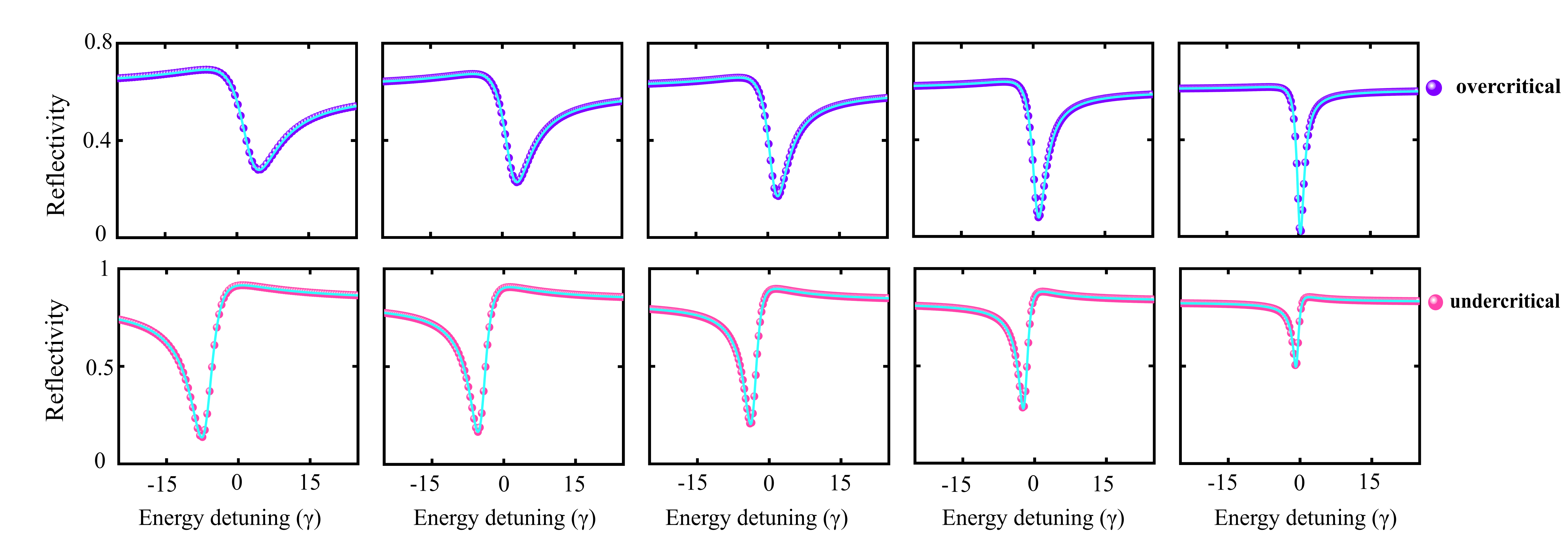}
	\renewcommand{\figurename}{Fig.}
	\caption{\label{Change_N} Fano resonances with different nuclear abundances at a fixed angle offset in the different regimes. Purple / pink dots are the numerical results calculated by CONUSS \cite{conuss-2000} at 2.38 / 2.28 mrad along with their fitted curves by the quantum model \cite{Heeg-2015-2,Heeg-2013} for overcritical and undercritical regimes respectively. The abundances of the nuclei are 100\%, 70\%, 50\%, 30\% and 10\% from left to right by turn.}
\end{figure*}

In the various cavity regimes, the relative phase of the two pathways is fixed at a distinct angle offset, and the Fano resonances changes with the different nuclear abundances as shown in Fig. \ref{Change_N}. It is clear that the asymmetry of the Fano resonances in each cavity regime does not change as nuclear abundance decreases due to the constant phase difference, while the narrower and the deeper or shallower profiles are the results of the varying CRS. As for the depth degree of the profile, it depends on the weight of the $[{\rm{Im}} (q)]^2$ in the Fano resonance \cite{Avrutsky-2013}. 
\begin{figure}[htbp]
	\centering
	\includegraphics[width=0.5\textwidth]{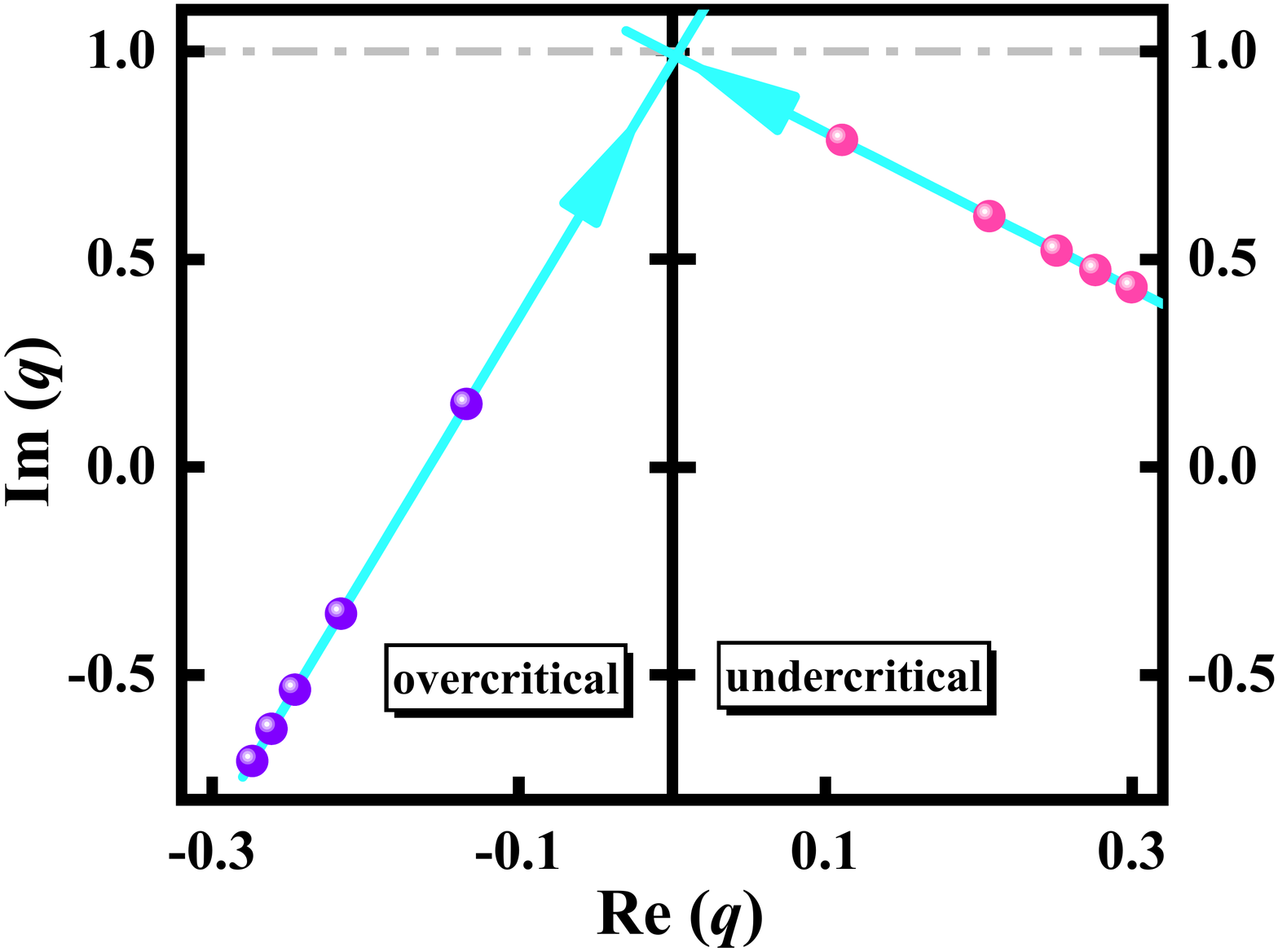}
	\renewcommand{\figurename}{Fig.}
	\caption{\label{Dissipation} The behaviors of the complex \emph{q} controlled by the nuclear abundance in different regimes. The fitted complex \emph{q} are shown by dots, and the direction of diminishing nuclear abundance is shown by the arrow.}
\end{figure}
The fitted complex \emph{q} from Fig. \ref{Change_N} are shown in Fig. \ref{Dissipation}.  As the nuclear abundance decreases, the CRS decreases correspondingly along the direction of the arrow. Remarkably, the trajectory of complex \emph{q} varies along the straight line with different slopes on the complex plane in both regimes, where the slopes depend on the relative phase $\varphi_{E}$ in different regimes. Furthermore, $\emph{q} = i$ is the cross point of different lines in the complex plane, where it corresponds to the minimal CRS in this work, viz. the bare cavity without collective nuclear resonance. It seems that there is an intuitive disagreement in contrast to the results in Ref. \cite{Andreas-2010}, where the behavior that the asymmetric parameter $q$ follows the straight line towards the dissipation-dominated point $q=i$ represents the resonant width is broadened by a stronger dissipation. However, in this work it corresponds to a narrower resonant width with the CRS gradually decreasing. This seemingly contradictory  phenomenon can be understood qualitatively from the perspective of interaction. The coupling strength between the cavity and the nuclear ensemble is proportional to $\sqrt{N}$, thus their energy exchange can be enhanced by the collective effect of the nuclear ensemble with increasing effective nuclear number $N$ \cite{Haroche-1983}, such that the dissipation process is suppressed equivalently in spite of the approximately contant cavity decay $\kappa$. 
As a result, the dissipation process suppressed by the collective effect among the nuclear ensemble, is different from the microwave cavity \cite{Andreas-2010}.

\begin{figure*}[!htbp]
	\centering
	\includegraphics[width=\textwidth]{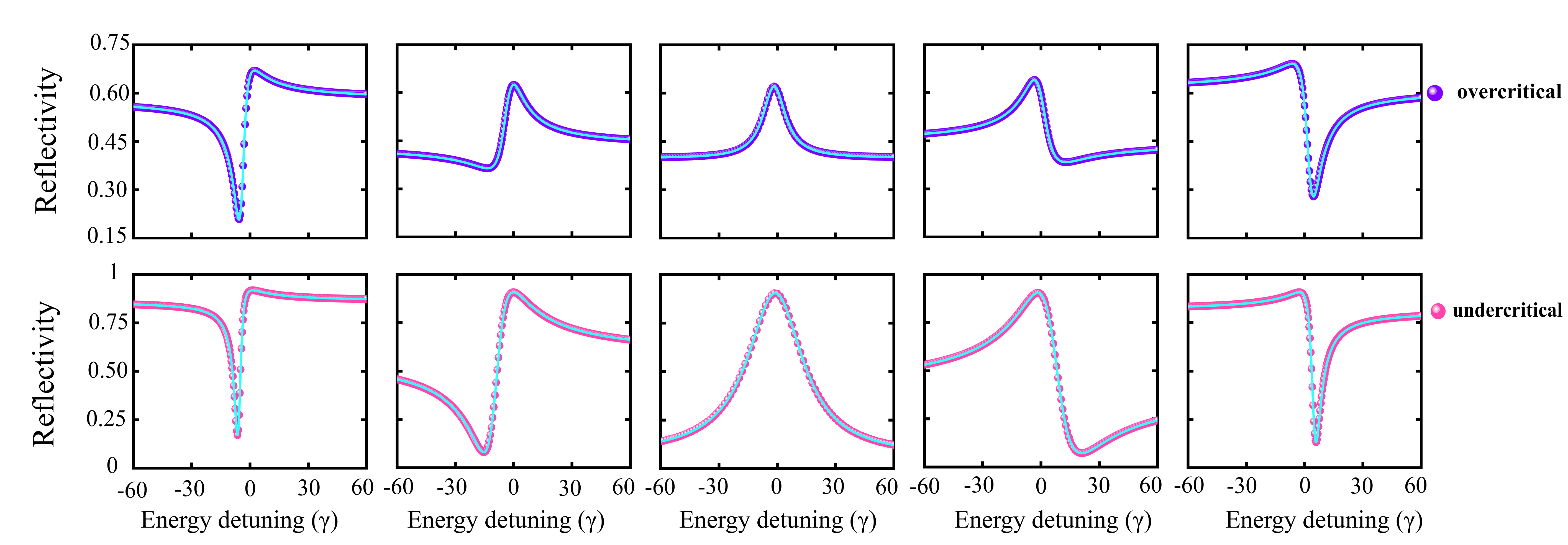}
	\renewcommand{\figurename}{Fig.}
	\caption{\label{Change_angle} Fano resonances at different angle offsets for overcritical and undercritial regime with 100\% nuclear abundance.  Purple / pink dots are the numerical results by CONUSS \cite{conuss-2000}  along with their fitted curves by the quantum model \cite{Heeg-2013,Heeg-2015-2} for overcritical and undercritical regimes respectively. The angle offsets are -50, -18, -2, 14 and 46 $\mu$rad from left to right by turn.}
\end{figure*}
\begin{figure*}[!htbp]
	\centering
	\includegraphics[width=0.8\textwidth]{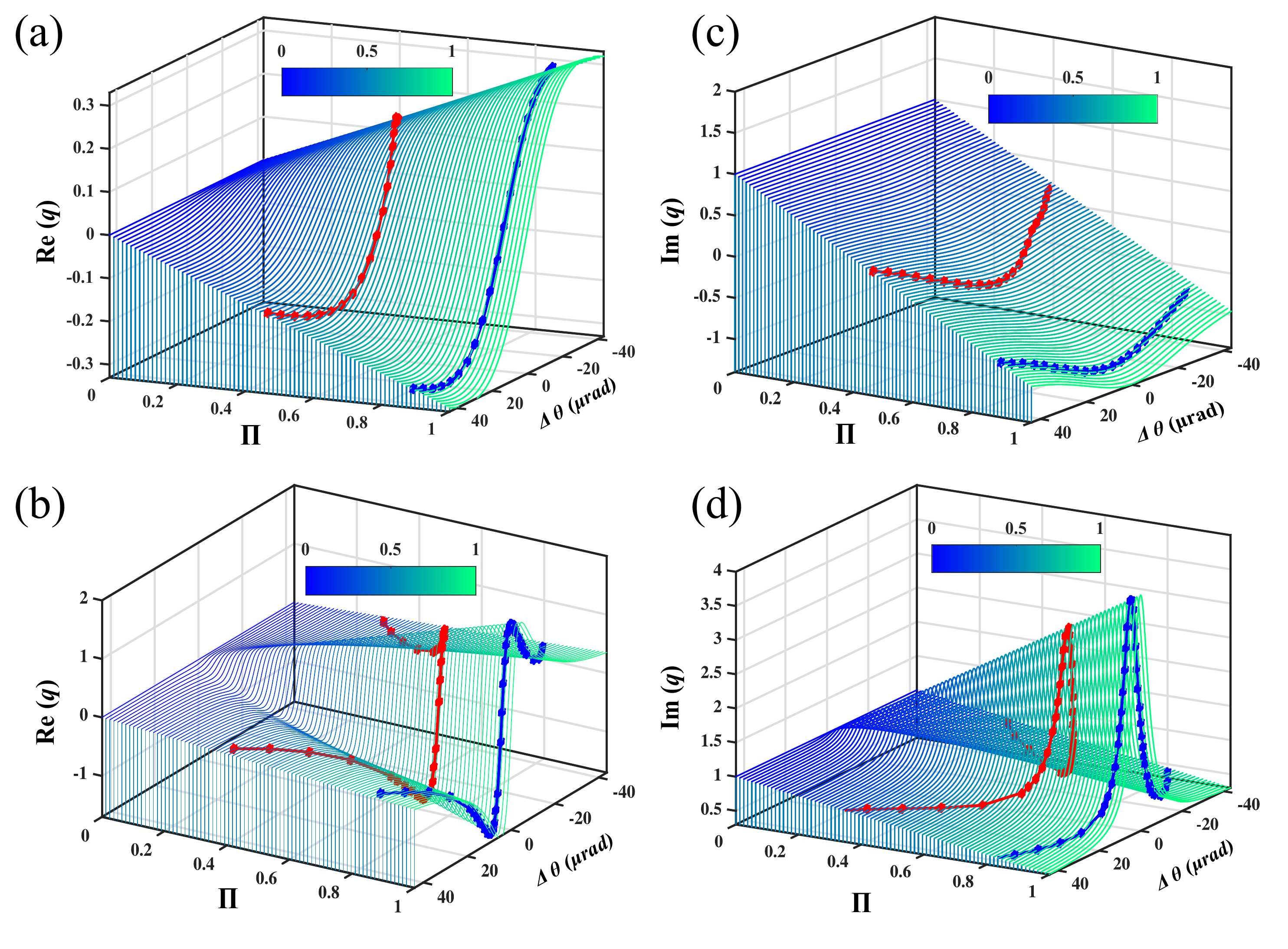}
	\renewcommand{\figurename}{Fig.}
	\caption{\label{angle-3Re-3Im} The dependences of the Re $q$ and Im $q$ on the CRS ($\Pi$) as well as the angle offsets $\Delta \theta=\theta - \theta_{1st}$. (a) and (b) are the curved surfaces of Re $q$ in the overcritical and undercritical regimes, whearas (c) and (d) are of the Im $q$ in different regimes correspondingly. The fitted complex $q$ with different nuclear abundance are represented by dot lines (red = 10\%, blue = 100\%).}
\end{figure*}

Furthermore, the CRS can also be manipulated conveniently by the cavity detuning in the thin-film planar cavity. In the overcritical and undercritical regimes, Fano resonances with different angle offsets are shown in Fig. \ref{Change_angle}. Obviously, the relative phase of the two pathways that determines the asymmetric shape of the Fano resonance changes with the variable angle offsets. Fig. \ref{angle-3Re-3Im} shows the fitted complex $q$ with various angle offsets under different nuclear abundances in more detail. Because the single atom broadening width $\Gamma_{\rm{s}}$ is constant at a fixed angle offset, the higher nuclear abundance leads to the stronger CRS. 
It is also noted that CRS is insensitive to the angle offset at a higher nuclear abundance. As for the behaviors of the Re ($q$) and Im ($q$), there are always negative Re ($q$) at the positive angle offset and positive Re ($q$) at the negtive angle offset, which is determined by the phase difference. The negative angle offset always corresponds to the phase range in $(-\pi/2,\pi/2)$, while the positive angle offset corresponds to the remaining range as shown in Fig. \ref{Fano_full}(c).
As for the valley and peak on the Im ($q$) curved surface for different regimes in Figs. \ref{angle-3Re-3Im}(c) and (d), it is  determined by the phase difference mainly. In the vicinity of the cavity mode, the relative phase $\varphi_E$ in the overcritical regime is around $-\pi/2$ while it is around $\pi/2$ in the undercritical regime, thus Im ($q$) takes its minimum and maximum values in the overcritical and undercritical regime, respectively.
\begin{figure}[htbp]
	\centering
	\includegraphics[width=0.5\textwidth]{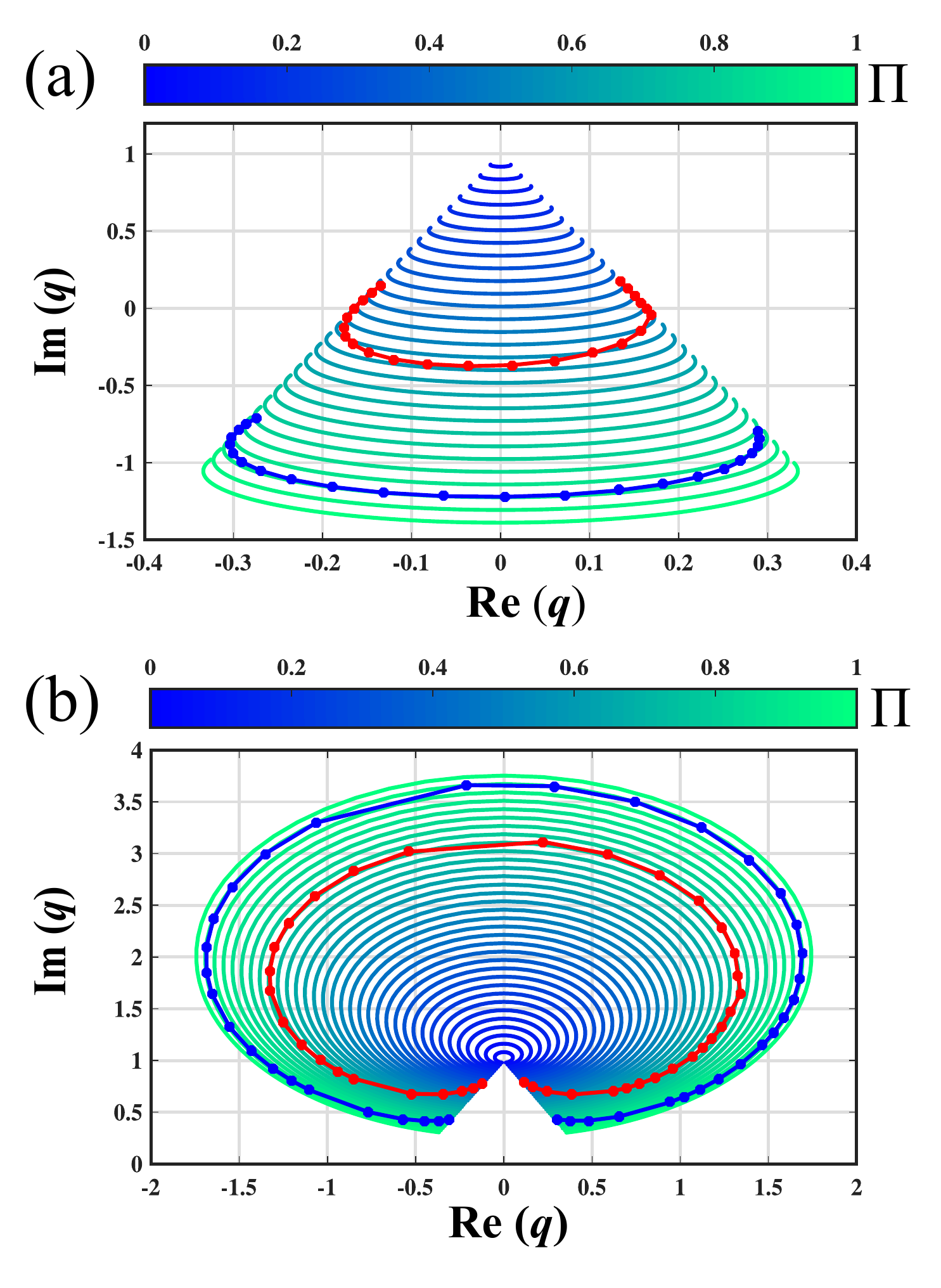}
	\renewcommand{\figurename}{Fig.}
	\caption{\label{3Re-Im} The behaviors of the complex $q$ controlled by the cavity detuning in the overcritical (a) and undercritical (b) regimes. Corlor bar represents the values of the CRS ($\Pi$). The red and blue dot lines are the fitted complex $q$ at different nuclear abundance (red=10\%, blue=100\%).}
\end{figure}
With the changeable angle offset, the behaviors of the asymmetric parameter \emph{q} in the complex plane are shown in Fig. \ref{3Re-Im}. Because the phase range is wider in the undercritical regime for the same nuclear abundance, more full arc trajectory is observed. At different nuclear abundance in the same regime, the radius of the arc trajectory shows visibly dependence on the nuclear abundance, i.e., a higher nuclear abundance corresponds to a larger radius.
In addition, since the CRS can also be adjusted by the angle offset, the trajectory in the complex plane is obviously bent towards the extreme point $q = i$ as the CRS decreasing at both ends. For the trajectories at different nuclear abundance shown in Fig. \ref{3Re-Im}, the degree of the distortion of the trajectory is significantly reduced owing to the narrower CRS range at a higher nuclear abundance.
The trajectory difference suggests that the radius of the arc trajectory mostly depends on the nuclear abundance, while the radius range is mainly affected by the angle offset.

\section{ Conclusion}
In summary, the dissipation process affected by the collective effect are studied via the Fano resonance in the thin-film planar cavity, where the asymmetric parameter $q$ is a complex value in the overcritical and undercritical regimes. In such systems, the resonant width is broadened by the collective effect of the nuclear ensemble, and the collective resonant strength mainly depends on the nuclear abundance, while its range can be regulated by the angle offset.   
For the Fano resonances with a stable phase difference at a fixed angle offset, the aysmmetric parameter $q$ show the straight lines as the nuclear abundance decreases, and the slopes are determined by the relative phase of the two pathways. The cross point $q = i$ of different lines corresponds to the dissipation-dominated case with the minimal CRS in the thin-film planar cavity where there is no energy exchange between the cavity and the nuclear ensemble. 
In addition, the CRS can also be manipulated by the angle offsets, and the asymmetric parameter $q$ presents the distorted arc trajectories correspondingly in the complex plane. As for the degree of the distortion, it is determined by the CRS range adjusted by the angle offset. Furthermore, the radius can be remarkably changed by the nuclear abundance. When the angle offset is modified at a higher nuclear abundance, the radius of the arc trajectory become noticeably larger, and the degree of the distortion of the trajectory is significantly reduced due to the narrower range of CRS.
The present work indicates that the strong collective effect of the nuclear ensemble not only can suppress the the dissipation process,
but also can change the radius of the trajectory. Recently, more complicated trajectories of the asymmetric parameter $q$ in the complex plane have also been observed, such as the circular trajectory in the single mode cavity as well as the cusps and loops induced by the mode overlap \cite{Lentrodt-2021}. These studies will attract more interest about the decoherence manipulation based on the thin-film planar cavity with flexible and controllable parameters space.

\section{Acknowledgments}
This work is supported by the National Natural Science Foundation of China (Grant No. U1932207), Strategic Priority Research Program of the Chinese Academy of Science (Grant No. XDB34000000), and the National Key Research and Development Program of China (Grant No. 2017YFA0402300). The financial support from the Heavy Ion Research Facility in Lanzhou (HIRFL) is also acknowledged.

\section{Appendix: Fitting process about the reflectivity} 
The reflectivity of the bare cavity simulated by CONUSS \cite{conuss-2000} can be seen in Fig. \ref{Fano_full}(b). However, there are two additional effects called heuristic extensions which need to be considered \cite{Heeg-2015-2}. The absorption from the bulk materials of the top layer would weaken the reflectivity intensity and also make a tiny dispersion.
Hence, the bare cavity $R_c$ can be fitted as follows \cite{Huang-2020},
\begin{align}
R_{c\rm{fit}} =&\;A \cdot ({-e^{i\varphi }} + \frac{{2{\kappa _R}}}{{\kappa  + i{\Delta _c}}}) \tag{A.1}
\end{align}
where A and $\varphi$ are used to consider the absorption effect and the dispersion of the top layer, respectively. In this work, the cavities have extremely tiny dispersion in order to get close to -1 as possible. The fitted cavity parameters such as cavity mode angle $\theta _{th}$, coupling strength $\kappa_R$ and the cavity decay rate $\kappa$ are shown in the Table \ref{fitting-parameters}. 

\begin{table}[!htb]
	\caption{\label{fitting-parameters} The fitted parameters in different regimes.}
	\centering
	\begin{tabular}{p{2.5cm}p{4cm}p{4cm}}
		\hline\hline
		&\multicolumn{1}{c}{overcritical} &\multicolumn{1}{c}{undercritical} \\
		\hline
		\multicolumn{1}{c}{A}&\multicolumn{1}{c}{0.77}&\multicolumn{1}{c}{0.94}\\
		\hline
		\multicolumn{1}{c}{$\varphi$}&\multicolumn{1}{c}{-0.02}&\multicolumn{1}{c}{0.038}\\
		\hline
		\multicolumn{1}{c}{$\theta_{1st}$[mrad]}&\multicolumn{1}{c}{2.338}&\multicolumn{1}{c}{2.320}\\
		\hline
		\multicolumn{1}{c}{$\kappa$[$10^{-2}\omega_0$]}&\multicolumn{1}{c}{1.938}&\multicolumn{1}{c}{0.7391}\\
		\hline
		\multicolumn{1}{c}{$\kappa_R$[$10^{-2}\omega_0$]}&\multicolumn{1}{c}{1.667}&\multicolumn{1}{c}{0.2711}\\
		\hline\hline
	\end{tabular}
\end{table}

In addition, owing to a low typical $Q$ value of the x-ray thin-film planar cavity, there is a tiny shift between the true mode angle $\theta_{th}$ and the accessible $\theta_{min}$ that corresponds to the minimum reflectivity for the material dispersion \cite{Huang-2021}. Hence, an extra energy shift should be considered for the Fano resonance, and the asymmetric profiles can be fitted by 
\begin{align}
|{R_{\rm{fit}}}{|^2} = a \cdot |R(\delta  + {\delta_{LS}}){|^2} + b \tag{A.2}
\end{align}
where $a$ is used to consider the absorption effect from the multilayer materials, $b$ is for the tiny background and $\delta$  considers the tiny energy shift.

\bibliography{bibDecoherence}
\end{document}